# *Transmission-Line Analysis of ε-Near-Zero (ENZ)-Filled Narrow Channels*

Andrea Alù[(1)], Mário G. Silveirinha[(1,2)], and Nader Engheta[(1)*]

[(1)]Department of Electrical and Systems Engineering, University of Pennsylvania

Philadelphia, PA 19104, U.S.A.

[(2)]Department of Electrical Engineering – Instituto de Telecomunicações,

Universidade de Coimbra, 3030 Coimbra, Portugal

E-mail: andreaal, msilveir, engheta @ee.upenn.edu

**Abstract**

Following our recent interest in metamaterial-based devices supporting resonant tunneling, energy squeezing and supercoupling through narrow waveguide channels and bends, here we analyze the fundamental physical mechanisms behind this phenomenon using a transmission-line model. These theoretical findings extend our theory, allowing us to take fully into account frequency dispersion and losses and revealing the substantial differences between this unique tunneling phenomenon and higher-frequency Fabry-Perot resonances. Moreover, they represent the foundations for other possibilities to realize tunneling through arbitrary waveguide bends, both in E and H planes of polarization, waveguide connections and sharp abruptions and to obtain analogous effects with geometries arguably simpler to realize.

---

[*] To whom correspondence should be addressed. E-mail: engheta@ee.upenn.edu

**1. Introduction**

The anomalous physics of artificial materials and metamaterials with index of refraction near zero has attracted the attention of physicists and engineers for decades [1]-[12]. Applications for directive radiation, enhancement of transmission, tailoring of the phase pattern and cloaking have been suggested over the years. As a further anomalous phenomenon associated with their properties, in a recent contribution by our group [11]-[12] it has been suggested that ultranarrow waveguide channels and bends may sustain a dramatic increase of transmission when filled with metamaterials with permittivity near zero. Heuristically, this is associated with the fact that the long wavelength in zero-permittivity materials allows transferring, in a "static-like" fashion, i.e., with essentially no phase delay, the field from input to output of a tight channel with arbitrary length, shape and geometry. More rigorously, it has been theoretically shown in [11] that the reflection from an arbitrarily shaped channel filled with a material with identically zero permittivity is surprisingly proportional to its longitudinal cross-sectional area. In other words, when such metamaterial fills a given channel connecting two waveguide sections, the tighter is the transition channel, the larger is the transmission through it, in principle independent of its specific shape.

Such "supercoupling" phenomenon has been recently demonstrated experimentally independently and almost simultaneously by two different groups. The experimental setup considered in [13] is based on a parallel-plate waveguide at microwave frequencies using a screen patterned with complementary split ring resonators to emulate the response of an ε-near-zero (ENZ) metamaterial that fills a narrow waveguide channel connecting two larger waveguide sections. Our arguably simpler experimental demonstration of this resonant tunneling phenomenon is based on a completely different conceptual approach, and exploits instead the intrinsic dispersion characteristic of a metallic waveguide near cut-off, as detailed in [12], [14], and briefly reviewed in this introduction.

For decades it has been well known how the intrinsic dispersion of a waveguide mode is electromagnetically analogous to the propagation in artificial materials and metamaterials. Rotman [15] has first introduced these concepts in order to realize artificial plasma-like materials at microwave frequencies using arrays of parallel plate waveguides and similar concepts have been recently employed to synthesize negative-index metamaterials [16]-[18]. In this sense, the intrinsic dispersion of the $TE_{10}$ mode in a rectangular metallic waveguide has been employed to realize an effective Drude-like negative permittivity, which has been used, together with split-ring resonators, to achieve negative-index propagation [16]-[18]. Similarly, the cloaking mechanism presented in [9], which requires in principle plasmonic materials, has been obtained using parallel-plate implants that emulate the response of a material with negative permittivity [19].

In a rectangular waveguide, the propagation of the dominant $TE_{10}$ mode may indeed behave equivalently to that of a transverse electromagnetic ($TEM$) wave traveling in a "material" with constitutive parameters [12], [15]:

$$\varepsilon_{eff} = \varepsilon_0 \left[ \varepsilon_r - c^2 / \left( 4 f^2 w^2 \right) \right]$$
$$\mu_{eff} = \mu_0 \tag{1}$$

where $\varepsilon_0, \mu_0$ are the free-space constitutive parameters, $\varepsilon_r$ is the relative permittivity of the dielectric filling the waveguide, $c = \left(\varepsilon_0 \mu_0\right)^{-1/2}$ is the velocity of light in free-space, $w$ is the waveguide width and an $e^{-i2\pi ft}$ time convention is assumed. It is seen, in particular, that at the cut-off frequency $f_0 = \dfrac{c}{2w\sqrt{\varepsilon}}$, $\varepsilon_{eff} = 0$. In other words, as far as the propagation constants are concerned, even using natural materials such as metals and standard dielectrics with positive values of permittivity, it is still possible to effectively achieve low or negative effective permittivity values by varying the width $w$ of a suitably designed rectangular waveguide [12].

In Ref. [14], we applied these concepts to demonstrate experimentally the supercoupling effect, and we have effectively realized the required zero-permittivity by simply using an ultranarrow rectangular channel filled with air. In this sense, we have proven experimentally that an analogous resonant tunneling may occur at microwave frequencies through a rectangular subwavelength channel operating at its cut-off frequency $f_0$, allowing dramatic squeezing of energy by all means equivalent to what is achievable in an analogous parallel-plate channel filled with ideal ENZ material. Consistent with the theory reported in [11]-[12], this tunneling is independent of the specific shape of the channel, even when bends or sharp abruptions are present along it. In [20], in fact, we have realized analogous experiments through narrow 90° and 180° bends, obtaining nearly-unitary transmission around the same frequency $f_0$.

It is worth underlining that in all our experiments the ENZ effect in the channels is not realized using subwavelength inclusions to tailor the material response of the channel, but rather with the suitable design of the lateral width $w$ of the waveguide, in order that the ultranarrow channel is operated at the cut-off frequency of the dominant $TE_{10}$ mode. After proper homogenization, a classic metamaterial and such waveguide section may both be treated as ENZ metamaterials. The achieved supercoupling and squeezing effect is fundamentally different from classic wave tunneling through waveguide channels or filters below cut-off [21]-[22], which relies on Fabry-Perot resonances. Quite differently, the enhancement of transmission made possible by ENZ materials is independent of the channel length, geometry and shape, and always occurs around the same frequency where $\varepsilon = 0$. Moreover, the field phase variation across the channel is interestingly uniform, consistent with the static-like property of ENZ propagation. In general, the theory of supercoupling has been developed and established in the ideal limit of $\varepsilon = 0$ [11]-[12], which due to frequency dispersion is an ideal condition achievable at a single frequency of operation, and which, in reality, may never be strictly observed in practice by the necessary presence of absorption. Here, applying familiar transmission-line concepts, we analyze in more detail this phenomenon considering frequency dispersion and a realistic deviation from ideal values of permittivities, providing a novel and distinct interpretation of the anomalous tunneling phenomenon and of its dependence over the geometry and the design parameters. The theoretical results derived here underline the drastic differences between this unique tunneling phenomenon and higher-frequency Fabry-Perot resonances. Moreover, they provide further insights into the phenomenon and suggest different possibilities and geometries to achieve analogous tunneling effect in arguably simpler or different geometries, which may be of interest for several purposes and applications, as analyzed and discussed in the following.

**2. Theoretical formulation: the U-shaped transition**

Consider the geometry of Fig. 1a, i.e., two identical parallel-plate (2D) waveguides of height $h$, filled with a material with (effective) permittivity $\varepsilon_{wg}$ and connected by a narrow channel of height $h_{ch} \ll h_{wg}$ and (effective) permittivity $\varepsilon_{ch}$. At the abruptions, two narrow transition channels of length $l_{ab}$ and permittivity $\varepsilon_{ch}$ have also been added, in order to accept the impinging wave and let it tunnel into the narrow channel, giving to the transition region uniformly filled with permittivity $\varepsilon_{ch}$, a characteristic U shape. All the material permeabilities are assumed equal to $\mu_0$. In [11] it was proven theoretically how, in the lossless limit for which $\varepsilon_{ch} = 0$, the reflection coefficient at the input of the transition region may be written in an elegant and exact closed form as:

$$R = \frac{ik_0\mu_0 A}{2h - ik_0\mu_0 A}, \tag{2}$$

where $k_0 = 2\pi f / c$ the wave number in free-space and $A$ is the total longitudinal cross-sectional area filled by the material with $\varepsilon_{ch} = 0$. It is noted how the result in (2) is an exact closed-form expression in this ideal limit and it does not depend on the specific shape of the channel. In particular, Eq. (2) counterintuitively implies that total transmission through the channel may be achieved for $h_{ch} \to 0$.

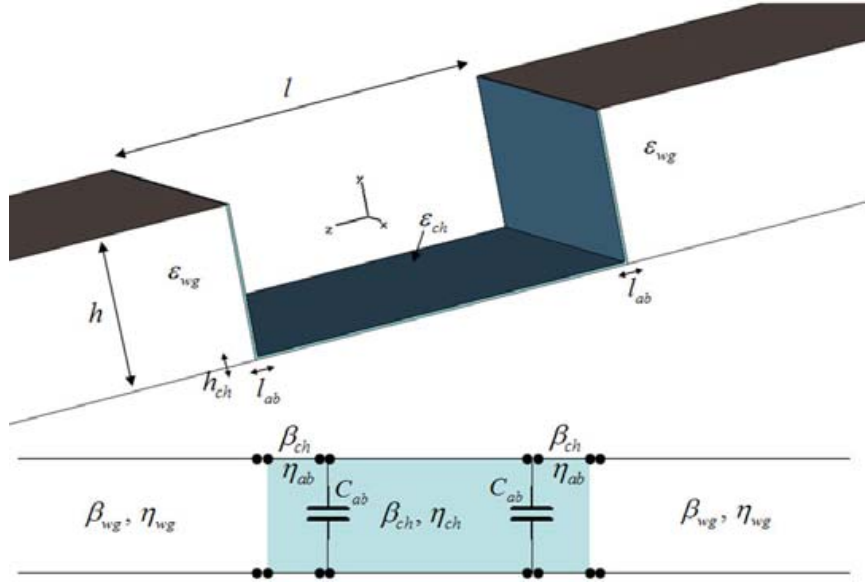

Figure 1 – (Color online) (a) Geometry of the problem: a narrow channel of height $h_{ch}$ and length $l$ connects two waveguide sections of height $h \gg h_{ch}$; (b) corresponding transmission-line model, in analogy with [13]. The structure is uniform along the $x$-direction.

In order to analyze the response of the structure of Fig. 1a in frequency and for values of $\varepsilon_{ch}$ different from zero, we may employ in this geometry an equivalent transmission-line (TL) model [23], as depicted in Fig. 1b. An analogous model has been used in [13] for a parallel-plate waveguide channel filled by ENZ materials. The outer waveguide sections may be described as TL segments with secondary parameters:

$$\beta_{wg} = 2\pi f \sqrt{\varepsilon_{wg} \mu_0}$$
$$\eta_{wg} = \frac{h}{w} \frac{2\pi f \mu_0}{\beta_{wg}} = \frac{h}{w} \sqrt{\frac{\mu_0}{\varepsilon_{wg}}},\quad (3)$$

being respectively the guided wave number and the normalized characteristic line impedance. The quantity $w$ is an arbitrary length (it may be considered unity) in the 2D geometry, and corresponds to the waveguide width in rectangular 3D geometries. $\eta_{wg} = V^+ / I^+$ is defined as the ratio between the voltage across the metallic plates and the current along the plates, when the TEM mode propagates along the $z$-direction. The characteristic line impedance $\eta_{wg}$ should not be confused with the wave impedance in the dielectric defined by $\sqrt{\mu_0 / \varepsilon_{wg}}$.

In the channel region, these quantities need to be redefined as $\beta_{ch}$ and $\eta_{ch}$ after the replacements $\varepsilon_{wg} \to \varepsilon_{ch}$, $h \to h_{ch}$, whereas in the transition channels (i.e., the arms of the U shape) the wave number is $\beta_{ch}$ and $\eta_{ab}$ is obtained from $\eta_{wg}$ replacing $\varepsilon_{wg} \to \varepsilon_{ch}$. It is noted that the line impedances are normalized to the waveguide height in each section, in order to ensure continuity of the voltage across the abruptions. The capacitive loads $C_{ab}$ in the TL model take into account the stored reactive fields at the abruption walls, associated with the evanescent modes excited at the abrupt discontinuities at the entrance and exit faces of the channel. For positive values of $\varepsilon$ in each region, $\beta_{wg}$ and $\eta_{wg}$ are real and positive quantities, consistent with propagation inside the waveguide. However, for negative values of $\varepsilon$ or when losses are considered, the proper square root branches should be chosen to have $\text{Im}[\beta] > 0$ and $\text{Im}[\eta] < 0$, consistent with the decay of evanescent waves and with the inductive properties of a waveguide filled with epsilon-negative materials [24]. In this way, the TL model in Fig. 1b may effectively describe in a compact and simple way the frequency dispersion of the waveguide transition in Fig. 1.

It is noted that this model effectively applies also to the 3D geometry considered in [14], consisting of a rectangular waveguide, in which the finite lateral width $w$ may modulate and determine the effective permittivity in each region, as discussed in the introduction and described by Eq. (1). In this case, in Eq. (3) and analogous formulas one should consider the effective permittivity $\varepsilon_{eff}$ in each region, and the value of $w$ in the formulas should be taken equal to the waveguide width.

The reflection at the entrance of the abruption region (area in blue) may be evaluated in closed-form for the model of Fig. 1b. In particular, the presence of the transition channels of length $l_{ab}$ is negligible in evaluating the reflection coefficient (their effect is important in reducing the effect of $C_{ab}$, as discussed in the next section), and we find:

$$R = \frac{2B\eta_{wg}^2\eta_{ch} - \left[\eta_{ch}^2 + \left(B^2\eta_{ch}^2 - 1\right)\eta_{wg}^2\right]\tan(\beta_{ch}l)}{\left[\eta_{ch}^2\left(B\eta_{wg} + i\right)^2 - \eta_{wg}^2\right]\tan(\beta_{ch}l) - 2\eta_{ch}\eta_{wg}\left(B\eta_{wg} + i\right)}, \qquad (4)$$

where $B = \omega C_{ab}$ is the load susceptance of the abruption sections (normalized to $w$). In the limit for which $B = 0$, i.e., the abruption effects are negligible, zero reflection may be achieved in two scenarios: (a) when $\tan(\beta_{ch}l) = 0$, which corresponds to a standard Fabry-Perot tunneling condition, strongly dependent on the length of the channel and on its geometry; (b) when $\eta_{ch} = \eta_{wg}$, i.e., when the two sections are impedance matched. The possibility that $\eta_{ch}$ may be equal to $\eta_{wg}$ may seem surprising at first glance, since the wave impedances $\sqrt{\mu/\varepsilon}$ in the two dielectrics are completely mismatched when the narrow channel is filled with an ENZ material. However, as discussed in the following, this does not necessarily imply that the line impedances are mismatched. In fact, quite interestingly, in case there is a strong cross-section mismatch between the waveguide sections and the channel, this second condition may be achievable when the (effective) permittivity of the channel region is sufficiently low to compensate the height mismatch. In particular, from (3) we obtain the simple condition:

$$\frac{h}{\sqrt{\varepsilon_{wg}}} = \frac{h_{ch}}{\sqrt{\varepsilon_{ch}}}, \qquad (5)$$

which would ensure line impedance matching and total transmission through the channel for $B = 0$. It is evident that when $h_{ch} \ll h$, it is required $\varepsilon_{ch} \ll \varepsilon_{wg}$, confirming Eq. (2) in the ideal limit of $\varepsilon_{ch} \to 0$. It should be noted, however, that, Eq. (2) applies to an arbitrary shape of the channel, even if it is valid only in the ideal limit $\varepsilon_{ch} \to 0$. Eq. (5), on the other hand, is a generalized condition valid also for finite values of permittivity of the channel, even though it is rigorously valid only for the straight channel of Fig. 1.

When considering the presence of the abruption loads, it is noted that a small finite value of $B$ slightly detunes the tunneling condition, but still allows a resonant tunneling basically independent on the channel length under the generalized condition:

$$\varepsilon_{ch} \simeq \frac{h_{ch}^2}{h^2}\varepsilon_{wg} - \frac{2B\sqrt{\mu_0}h_{ch}}{k_0 lw} + \frac{\mu_0 B^2 h_{ch}^2}{w^2}, \tag{6}$$

which has been obtained under the approximation $\tan(\beta_{in}l) \simeq \beta_{in}l$, valid around the ENZ operation of the channel of interest here. It is noted that Eq. (6) implies again $\varepsilon_{ch} \to 0$ when $h_{ch} \to 0$. In other words, despite the presence of a finite abruption admittance due to the height mismatch, resonant tunneling and total transmission are achievable around the ENZ operation of the channel, provided that its cross-sectional area is sufficiently small, supplying an interesting transmission line interpretation to the results presented in [11]. It is noted that Eq. (6) establishes a condition on the (effective permittivity) $\varepsilon_{ch}$ for obtaining *total* transmission, resonant tunneling and supercoupling. For non-zero values of $h_{ch}$ and $B$, the predicted value of $\varepsilon_{ch}$ is distinct from zero (but close to it), implying that at the (different) frequency for which $\varepsilon_{ch} = 0$ the transmission is not ideally unity, despite being possibly high. This is consistent and provides a generalization of the ideal result represented by (2), which is limited to the single frequency for which $\varepsilon_{ch}$ is identically zero. In other words, Eq. (2) does not pose a minimum limit to the reflection from the channel, but only evaluates the reflection at the frequency for which $\varepsilon_{ch} = 0$. At a close frequency, for which Eq. (6) is verified, the reflection may become identically zero despite the finiteness of $h_{ch}$ and $B$. It is worth underlining, moreover, that since the values of $B$ are usually relatively small and positive, Eq. (6) suggests that total tunneling may occur under some conditions (mainly when $B$ is not negligible, which happens when the transition regions are not considered, or not large enough) for low negative values of $\varepsilon_{ch}$, implying that the weakly inductive properties of the channel slightly below cut-off may compensate the small capacitive impedance of the abruption.

It is evident how the supercoupling phenomenon is very distinct from a Fabry-Perot tunneling, being totally independent of the length of the channel and being based on the anomalous line impedance match ($\eta_{ch} = \eta_{wg}$) provided by the ENZ material. In a 3D rectangular waveguide the same results may be obtained operating at the channel cut-off frequency, for which effectively $\varepsilon_{ch} \to 0$, as discussed above and experimentally verified in [14] and [20].

In Figure 2, as an example, we have considered a rectangular waveguide with uniform width $w = 2h = 10.16\,cm$ and $l_{ab} = h_{ch} = h/64 = 0.8\,mm$, consistent with the experimental setup analyzed in [14] and the geometry of Fig. 1. The outer waveguide sections are filled with Teflon (with permittivity $2\varepsilon_0$) and the ultranarrow channel is filled with air (with permittivity $\varepsilon_0$). Two different channel lengths are considered: $l = 12.7\,cm$ (black lines) and $l = 10.2\,cm$ (red lines).

The cut-off frequency of the channel, for which effectively $\varepsilon_{ch} = 0$, in this geometry arises at $f_0 = 1.47638\,GHz$, at which $\varepsilon_{wg} = \varepsilon_0$. Neglecting the presence of $C_{ab}$, Eq. (5) predicts resonant tunneling at $f = 1.47674\,GHz$, slightly above $f_0$, due to the finite value of $h_{ch}/h$. This is consistent with the dashed lines in Fig. 2, which reports the power transmission coefficient through the channel, as predicted by TL theory neglecting the presence of $C_{ab}$. Full-wave simulations obtained using CST Microwave Studio [25] confirm the possibility of resonant transmission around the cut-off frequency, but slightly below the value predicted by TL theory, as reported by the solid lines in Fig. 2, at frequency $f = 1.4562\, GHz$. In order to match this transmission frequency, it is necessary to consider a value of capacitance $C_{ab} = 1.6\, pF$ in Eq. (4). This is consistent with the negative shift in $\varepsilon_{ch}$ produced by a small positive $B$ in Eq. (6). The dashed lines correspond to the curves yielded by the TL model considering this value of $C_{ab}$, showing how perfect agreement at the tunneling frequency is obtained in this case between full-wave simulations and TL theory. It is noticeable how the resonant frequency in this scenario does not depend on the length of the channel, but it is simply based on the impedance matching ensured by the ENZ response of the channel, irrelevant of its geometry and shape.

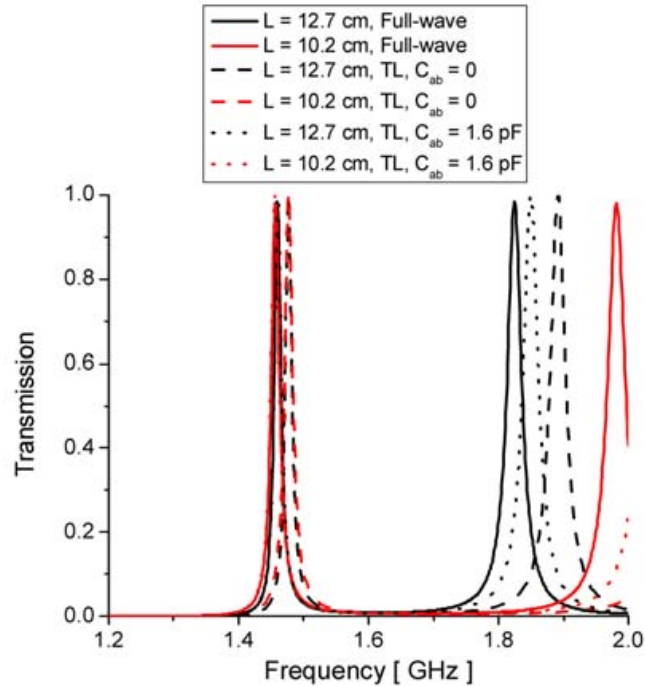

Figure 2 – (Color online) Power transmission through a narrow channel in a rectangular waveguide as in Fig. 1, with $l_{ab} = h_{ch} = 0.8\,mm$, $w = 2h = 10.16\,cm$. The outer sections are filled with a material with $2\varepsilon_0$, whereas the transition region has permittivity $\varepsilon_0$. $l$ is varied as indicated in the legend. Solid lines refer to full-wave simulations performed with [25], dashed lines use the TL model with $C_{ab} = 0$, dotted lines use $C_{ab} = 1.6\,pF$.

At larger frequencies, another tunneling peak is visible, due to a classic Fabry-Perot resonance. This is obtained near the frequency for which $\tan(\beta_{ch} l) = 0$, consistent with Eq. (4). It is evident how this resonance, however, is strongly dependent on the length of the channel, and it is very different from the supercoupling operation. Also at this frequency the presence of $C_{ab}$ allows a better matching with full-wave simulations, even if a larger value of $C_{ab}$ should be considered for a perfect agreement, consistent with the increase of the abruption capacitance with frequency [23].

Another remarkable difference between classic Fabry-Perot resonances and the supercoupling effect resides in the nearly uniform phase and amplitude distribution across the channel, independent of its total length, due to the quasi-static properties of the ENZ region. This is depicted in Fig. 3, which compares the phase of the magnetic field (Fig. 3a-b) and the electric field distributions (Fig. 3c-d, snapshot in time) along the channel considered in Fig. 2 with $l = 12.7\,cm$ at the two tunneling frequencies $f_0 \simeq 1.46\,GHz$ (supercoupling effect, Fig. 3a,c) and $f \simeq 1.82\,GHz$ (first Fabry-Perot resonance, Fig. 3b,d). Drastic differences are evident in the two scenarios: the supercoupling phenomenon, in fact, does not rely on an intrinsic resonance of the channel, but rather on impedance match and energy squeeze, which provides uniform electric field distribution (much enhanced across the channel due to the squeezing) and low phase delay. In comparison, the magnetic field amplitude inside the channel (not reported here) is comparable with the impinging one [12]. On the other hand, the Fabry-Perot resonance is characterized by a standing-wave distribution, with strong variations of amplitude and phase along the channel, such that both the electric and magnetic fields are enhanced. The small lowering of the tunneling frequency due to the reactive fields at the abruption, as described above, slightly affects the uniform-phase property across the channel (since the channel does not possess an identically zero permittivity, but a slightly negative one). Due to the presence of weakly evanescent impinging and reflected modes inside the channel, a small phase delay across the channel is produced by their interference. For the geometries of Fig. 2, the phase delay between entrance and exit faces is limited to $4°$ and $5°$, for $l = 12.7\,cm$ and $l = 10.2\,cm$ respectively. This is consistent with the results in [11]-[14], corresponding in this case to more than 95% reduction of the effective phase delay in the transition region. In comparison, the first Fabry-Perot resonance provides a $180°$ phase shift, consistent with Fig. 3.

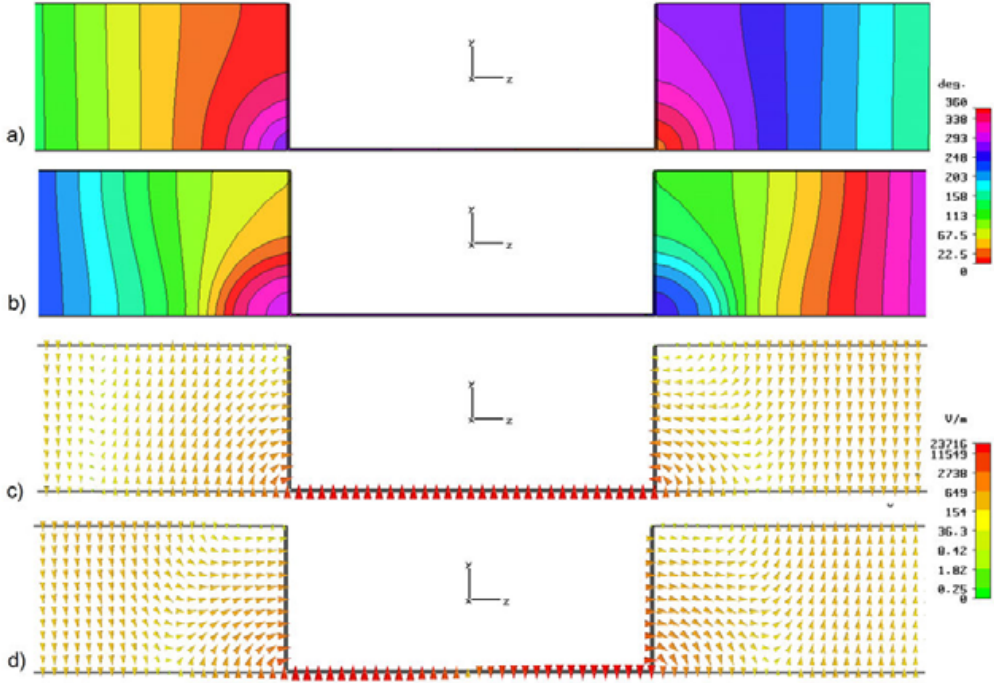

Figure 3 – (Color online) (a, b) Phase of the magnetic field and (c,d) electric field distribution (snapshot in time) for the geometry of Fig. 2 with $l = 12.7\,cm$ at the two tunneling frequencies $f_0 \simeq 1.46\,GHz$ (a, c: supercoupling effect) and $f \simeq 1.82\,GHz$ (b, d: Fabry-Perot resonance).

The supercoupling and energy squeezing produced at the ENZ frequency of operation may have several appealing applications, i.e., perfect coupling of distant waveguides, filtering independent of the length of the channel, and sensing applications that may benefit from the strong and uniform electric field induced across the channel region.

## 3. Squeezing energy into an ultranarrow waveguide: relevance of the transition channels

Following the results of the previous section, it is evident how the supercoupling phenomenon is based on the line impedance matching of an ultranarrow channel having near zero (effective) permittivity with an outer waveguide of much larger cross section, very different from the intrinsic resonance of the channel length associated with a standard Fabry-Perot resonance. Therefore, we may foresee the possibility of squeezing the impinging energy inside an ultranarrow channel even if there is no exit side to the channel, for example if the ENZ channel is either infinitely extended or terminated by a matched load. This effect highlights the drastic difference with a classic Fabry-Perot tunneling phenomenon, which indeed requires a standing-wave resonance, and therefore an entrance and an exit side. This property may suggest exciting potential applications in which the energy squeezed inside the channel may be absorbed or used for purposes different than tunneling. Such energy absorption would indeed smother any Fabry-Perot resonance based on multiple reflections along the channel length.

Considering only the first transition in the geometry of Fig. 1, as depicted in the inset of Fig. 4, the reflection coefficient at the entrance port is given by the following formula:

$$R = \frac{\eta_{ch} - \eta_{wg} + iB\eta_{ch}\eta_{wg}}{\eta_{ch} + \eta_{wg} - iB\eta_{ch}\eta_{wg}}, \qquad (7)$$

where once again the irrelevant small transition channel (i.e., the arms of the U channel) at the entrance plane has been neglected in the calculation for simplicity. The condition of minimum reflection is achieved when Eq. (5) is verified, i.e., near the ENZ operation of the channel for which $\eta_{ch} = \eta_{wg}$, providing a minimum reflection $R_{min} = -\dfrac{B\eta_{wg}}{2i + B\eta_{wg}}$. Of course in this scenario no Fabry-Perot resonance may be supported. In this case the presence of a finite abruption admittance $B$ cannot be compensated by slightly detuning the supercoupling frequency; as for a finite channel length, a non-zero reflection is always expected for finite values of $B$. This phenomenon actually shows the importance of the presence of the transition channels, as considered in [11]. In the following, we analyze their effect quantitatively.

The quasi-static equivalent susceptance of an abrupt transition like the one of interest here, uniformly filled by a material with permittivity $\varepsilon$, may be evaluated using variational methods [23], [26]. Its value for the case at hand is given by the following formula [26]:

$$B = \frac{k_0 h \sqrt{\varepsilon}}{\pi} \left[ 2 \ln \frac{h^2 - h_{ch}^2}{4 h h_{ch}} + \left( \frac{h}{h_{ch}} + \frac{h_{ch}}{h} \right) \ln \frac{h + h_{ch}}{h - h_{ch}} \right], \tag{8}$$

which shows how indeed a large cross-section mismatch, i.e., $h_{ch} \ll h$, as expected, considerably increases the abruption admittance $B$. For $h_{ch} \to 0$, the value of $B$ would tend to infinity, unless the material filling the abruption region has identically-zero permittivity. This implies that the presence of a uniform material surrounding the abruption with a low permittivity value may sensibly lower the effects of the abruption, due to its static-like properties discussed above. This explains the fundamental role taken by the transition channels, in the limit $\varepsilon_{ch} \to 0$ considered in [11], in lowering the value of $B$ for such sharp abruptions. Here the transition channels may play an equally important role in maximizing the transmission through the abruption. Clearly, the length $l_{ab}$ of the transition channels inside the waveguide section should be large enough to ensure capturing most of the evanescent modes stored at the abruption, in such a way that Eq. (8) is applicable, but on the other hand it should remain electrically small, not to create unwanted extra reflection. In the limit of $\varepsilon_{ch} \simeq 0$, which is of interest here, both these conditions are satisfied with small values of $l_{ab}$, since the evanescent fields are very much attenuated and concentrated by the static-like wave properties in the ENZ region, whereas the electrical length of the transition channels is reduced by the corresponding long wavelength.

Figure 4, as an example, shows the power transmission at the entrance abruption of Fig. 1, evaluated using [25], considering that the narrow channel is now impedance matched at its end (as if it were ideally extended to infinity). The different curves correspond to different values of $l_{ab}$ and the geometry corresponds to the rectangular waveguide of Fig. 2, being the ENZ response once again obtained with the presence of a finite width $w$. It is evident how increasing the length of the transition channel, may enable maximal energy squeezing and impedance matching between the two waveguide sections, obtaining nearly unity transmission and zero reflection despite the huge cross-section mismatch ($h_{ch} = h/64$). Already with $l_{ab} = 4h_{ch}$ reflection at the abruption and the associated admittance are close to zero. This effect clearly highlights the difference between this phenomenon and any other Fabry-Perot tunneling effect, which may not be supported over a single abruption, as confirmed by the plot, and therefore it confirms the distinct features of the supercoupling phenomenon.

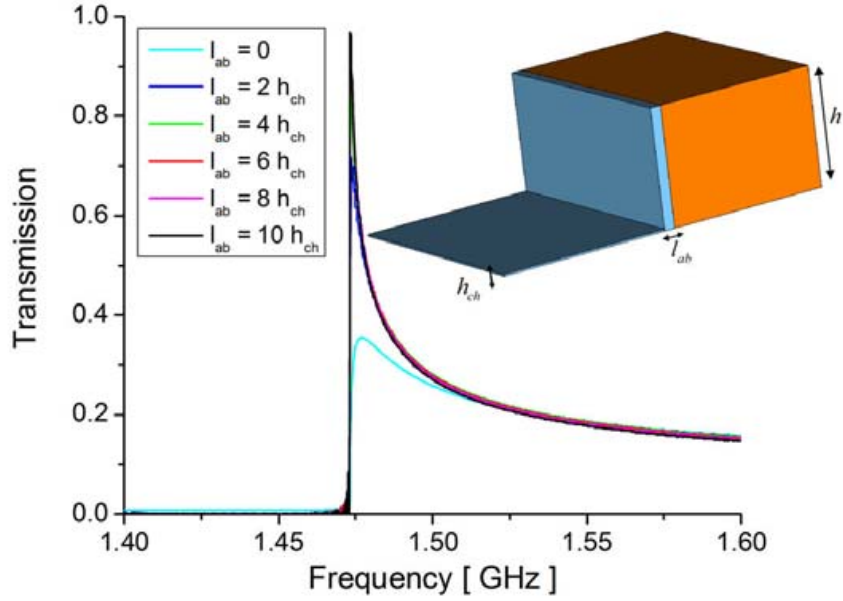

Figure 4 – (Color online) Power transmission evaluated using [25] through a single abruption in a rectangular waveguide as in Fig. 2, with $h_{ch} = 0.8\,mm$, $w = 2h = 10.16\,cm$. The outer section (orange) is filled with a material with $2\varepsilon_0$, whereas the ultranarrow channel (blue) has permittivity $\varepsilon_0$. $l_{ab}$ is varied as indicated in the legend.

The figure shows that without the presence of a transition channel ($l_{ab} = 0$), maximum transmission is found at the cut-off frequency $f_0$ of the narrow channel, but its peak is less than $40\%$, due to the presence of the abruption admittance. However, introducing the transition channels and increasing $l_{ab}$ it is possible to dramatically increase the energy squeeze inside the narrow waveguide, up to total transmission and supercoupling in the narrow waveguide, despite the abruption and the absence of reflections at the end of the channel. The energy in the channel may now be absorbed and used for different purposes without affecting the supercoupling properties of the transition region, since the transmission enhancement is not based on the presence of the exit face of the channel. Such operation is not possible with Fabry-Perot resonant tunneling which requires the formation of a standing wave. Using Eq. (7), the capacitance $C_{ab}$ associated with the abruption when $l_{ab} = 0$ may be evaluated as equal to $4\,pF$. Its value may be dramatically reduced considering the presence of the transition channel, as shown in the curves of Fig. 4.

It is worth noticing how, consistent with our previous theory, the tunneling frequency is not affected by the variation of $l_{ab}$, being simply related to the ENZ properties of the narrow waveguide. Once again, the energy squeezing achieved in this geometry has been obtained without relying on any subwavelength inclusion to form the required ENZ response in the narrow waveguide section, but rather is based on the proper choice of the width $w$ of the rectangular waveguide and of the two different materials, Teflon and free-space, filling the two waveguide sections, respectively. These results may be of interest in applications that require squeezing the electromagnetic energy from a larger conduit to an ultranarrow channel, in order to facilitate the use and absorption of the impinging power for various purposes.

The presence of the transition channels is of extreme importance in this scenario in order to achieve maximized (total) transmission into the ultranarrow waveguide by effectively canceling the value of the abruption admittance $B$. In the examples of the previous section where the channel had also an exit face, the presence of the transition channel is not strictly necessary, since supercoupling is still possible and a finite $B$ has the effect of just slightly detuning the tunneling frequency. As an example, Fig. 5 shows the transmission in amplitude and phase for a parallel-plate waveguide as in Fig. 1 (same geometry as in Fig. 3, but in the 2D case) filled with an ideal Drude material following the same dispersion as predicted by Eq. (1) and for a rectangular waveguide as in Fig. 3. The two geometries are simulated both without transition channels and with $l_{ab} = h_{ch}$. In Fig. 5a the thin solid line represents the dispersion of the (effective) $\varepsilon_{ch}$ with frequency, which crosses zero at the cut-off frequency $f_0$.

It is noticed how the behavior of the different waveguides is similar in the 2D and 3D geometries, consistent with the results of [12], and how the presence of the transition channels affects only in a minor way the supercoupling frequency in both cases. When the transition is not present, parallel-plate and rectangular waveguide have very similar properties, supporting tunneling at a frequency for which $\varepsilon_{ch}$ is negative and close to zero, as predicted by Eq. (6). When the channels are added, their effect shifts up the supercoupling frequency, closer to $f_0$. The residual value of $B$ is slightly larger in the rectangular geometry, for which a somewhat larger $l_{ab}$ would be required to tune the supercoupling frequency exactly at $f_0$, due to the larger reactive fields at the rectangular abruption. Anyhow, Fig. 5 supports the complete correspondence between the 2D parallel-plate problem filled with an ideal Drude metamaterial and the rectangular geometry that we have described above, filled with air.

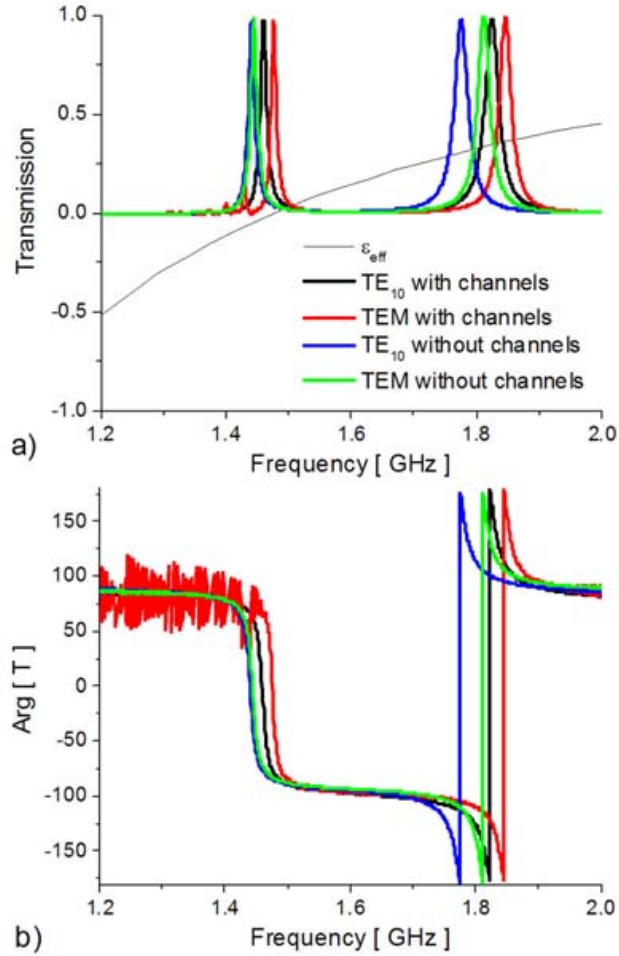

Figure 5 – (Color online) (a) Power transmission and (b) phase of the transmission coefficient evaluated with [25] through the channel of Fig. 3, comparing the cases of a parallel plate waveguide filled by an ideal Drude metamaterial and a rectangular waveguide as in Fig. 3, for which the material dispersion is provided by the waveguide width $w$. In the case transition channels are used their thickness is $l_{ab} = h_{ch}$. The thin gray line in panel (a) indicates the variation of $\varepsilon_{ch} / \varepsilon_0$ with frequency.

The phase of the transmission coefficient in Fig. 5b confirms the static-like property of the ENZ channel in both configurations, for which, despite its length a total phase delay of just few degrees is measured across the channel (compared to the 214° expected if the channel were filled with the same material as the outer waveguide section).

As an aside, it is noted that the drastic difference between the supercoupling phenomenon and a classic Fabry-Perot tunneling may make the former much more robust to the presence of losses. In fact, as proven in this section, the suitable design of the transition channels may make the entrance abruption completely matched with the outer waveguide sections, independent of the presence of the exit face of the channel, despite a huge cross-section mismatch. This implies that possible presence of losses or imperfections inside the channel may not influence the zero reflection at its entrance. Still the transmission may be affected by these losses, due to the large electric field necessarily induced inside the channel, but the reflection at the abruption would remain minimized. The higher-order Fabry-Perot resonances, on the other hand, requiring a strong standing-wave contribution for their sustainability, would be much more sensitive to the presence of losses, possible absorption or other imperfections.

To conclude this section, it is interesting to note that Eq. (8) implies, as an interesting corollary, that any abruption, discontinuity, sharp bend or geometry/shape modification in the ENZ channel at the supercoupling frequency would not sensibly affect its tunneling properties. This is because the associated propagation at the supercoupling frequency inside the narrow channel is static-like in nature, implying total transmission and negligible associated evanescent fields and abruption admittance at any of these abruptions. This is consistent with the results in [11], which are valid independent of the specific geometry and shape of the channel in the limit of $\varepsilon_{ch} = 0$.

**4. Alternative geometries: shrinking and bending in the H plane**

Having established a general theory for interpreting the supercoupling phenomenon in the geometries analyzed in [11]-[14], in this section we can apply these concepts to the design of alternative setups that may have interesting potentials for various applications.

As a first example, we notice from Eq. (1) that by modulating the width $w$ of a rectangular waveguide at different sections, it may be possible to effectively realize different permittivity profiles, even in waveguide sections filled with the same material. This may become particularly interesting for the easiness of realization of the supercoupling phenomenon, which may avoid the use of different materials to fill the different waveguide sections involved. For instance, in the previous sections we have considered rectangular waveguides having the same uniform widths in each of their sections, and we have selected different materials with different permittivities (i.e., Teflon and air) to fill each region in order to achieve the required effective permittivity. If the narrow channel was designed to be at cut-off when filled by air at frequency $f_0$, the rest of the waveguide was required to support a propagating mode at the same frequency, and therefore it was filled by Teflon, which has a larger permittivity. Here, on the other hand, we consider using the same material for all the sections (for simplicity just air), but we modulate the width in the channel region to achieve a similar variation in the effective permittivity, following Eq. (1).

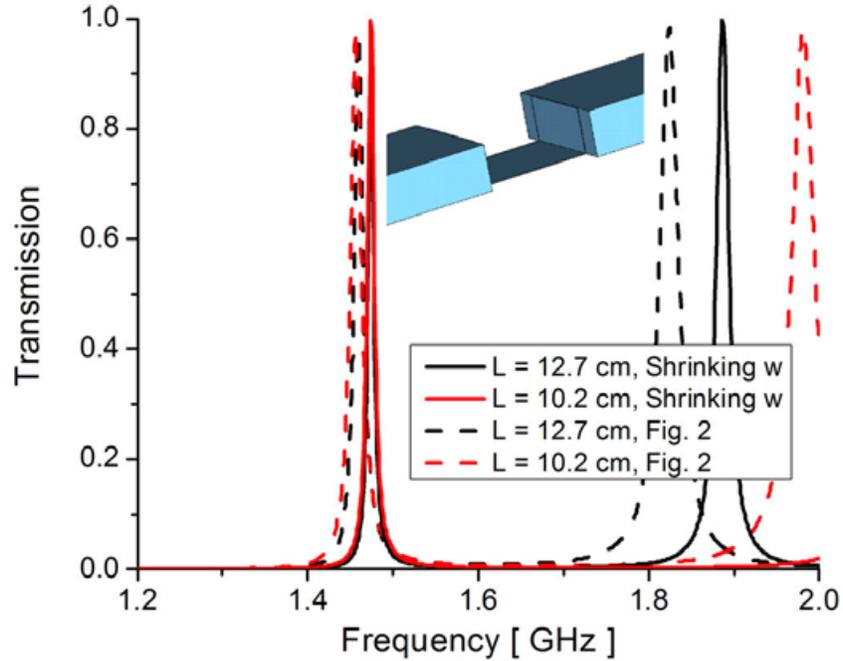

Figure 6 – (Color online) Power transmission, evaluated using [25], through a similar geometry as in Fig. 2, but varying the width $w$ in the different waveguide sections instead of varying the material filling the waveguide, as depicted in the inset.

In Figure 6, we have reported our full-wave simulation results for the abruption geometry designed following this concept, as depicted in the inset of the figure. The chosen channel lengths, its height and the length of the transition channels are the same as for the geometry in Fig. 2, even if now the whole waveguide is filled with air with permittivity $\varepsilon_0$. To be "electromagnetically" equivalent to the problem of Fig. 2, following Eq. (1) we have increased the width of the outer waveguide sections to $w = 15\,cm$ and we have compared the results for the two different channel lengths with those in Fig. 2.

It can be clearly seen how the effects are very similar, achieving a supercoupling transmission around the frequency $f_0$, independent of the channel length. It is interesting to observe how in this situation the tunneling frequencies are closer to the ideal values predicted by TL theory that neglect the abruption loads, or in other words, how the abruption admittance in this geometry is interestingly lower than in the geometry in Fig. 2. This is easily explained by the fact that an abruption in the H plane, as the one introduced by the width variation, is well known to provide an inductive admittance, which is summed algebraically to the capacitive one due to the sharp E plane height abruption. The two effects (one capacitive and one inductive) compensate each other in this geometry, bringing the full-wave simulations closer to the ideal behavior of the supercoupler with no abruption admittance. A similar trend is verified also for the higher frequency Fabry-Perot resonances in Fig. 6, which, as expected, depend strongly on the channel length and geometry.

As a second important application that follows the findings of the previous sections, it may be possible to consider arbitrary bends and abruptions along the channel without sensibly affecting the supercoupling mechanism. If bending in the E plane has been proven theoretically [12] and experimentally [14], [20] in recent works, here we want to show that a similar result may be obtained even for arguably more arduous H plane bends. Bending in this plane may occur only for 3D geometries, and maintaining the ENZ properties of the channel requires keeping the waveguide width constant in the bending process. However, the robustness of the ENZ operation, as we describe in the following, provides sufficient flexibility in the design.

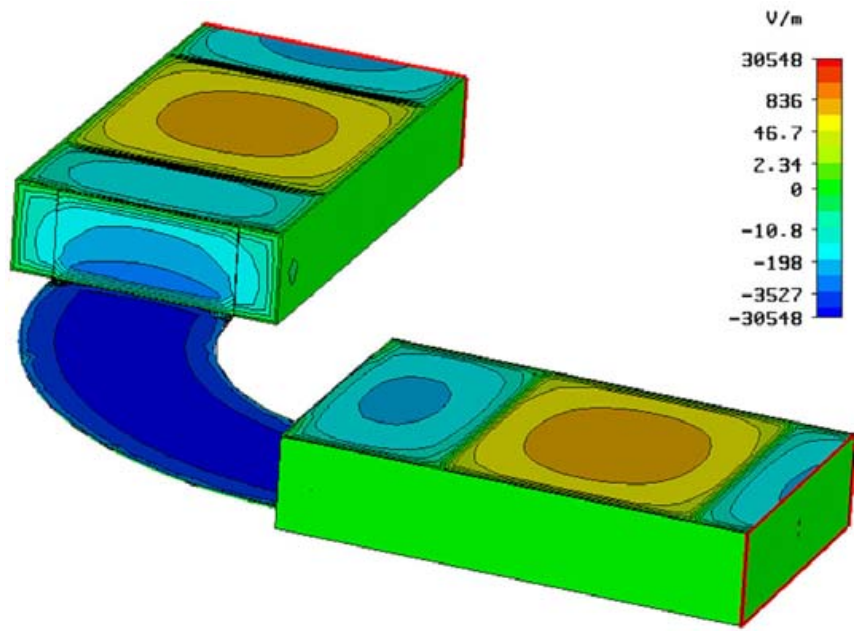

Figure 7 – (Color online) Geometry and electric field distribution (snapshot in time, normal component) through a supercoupler with a 90° bend in the H plane. The average rotation radius in this case is $R = 17.8\,cm$. The corresponding time-domain animation has been reported in EPAPS [27].

Consider the geometry of Fig. 7, which consists of similar waveguide geometry as in Fig. 6, i.e., rectangular waveguides with width $w = 15\,cm$ and height $h = 5.08\,cm$, connected through an ultranarrow channel of height $h_{ch} = h/64 = 0.8\,mm$ bending by 90° in the H plane. All the waveguide sections are filled by air, and therefore we have employed the technique of shrinking the H plane width, as described above, to achieve ENZ operation of the channel at frequency $f_0$, having a smaller width $w_{ch} = 10.16\,cm$, consistent with Fig. 6. The channel rotates in the H plane with an average radius of curvature $R = 17.8\,cm$, maintaining the same width all over the bend. Proper transition channels are also considered at the E plane abruptions, consistent with the previous discussion, with $l_{ab} = h_{ch}$.

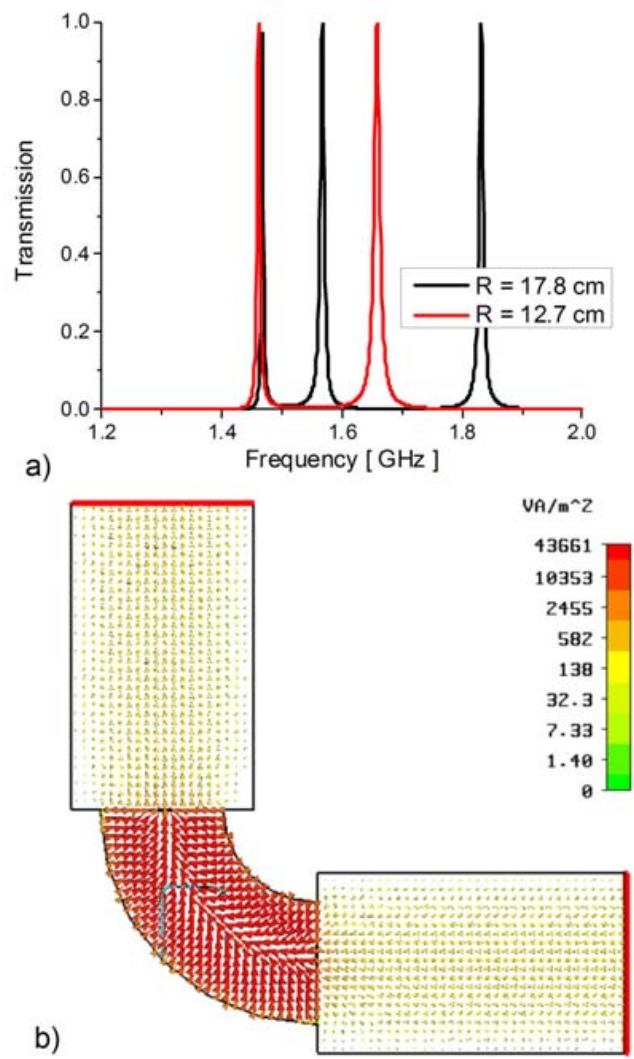

Figure 8 – (Color online) (a) Power transmission for the geometry of Fig. 7; (b) Real part of Poynting vector distribution on the H plane with $R = 12.7\,cm$ at the cut-off frequency $f_0$.

Figure 8a reports the transmission coefficients through the bend for two different radii of curvature, as simulated with [25]. Notice that, despite the channel length, which in this case is relatively long, the huge cross-section mismatch and the bending in the H plane, total transmission is achieved, independent of the length of the channel and its radius of curvature, at the ENZ operation frequency $f_0$. Other Fabry-Perot resonant peaks are visible also in this scenario at higher frequencies, even though they are very much dependent on the length of the channel and on its shape. For what discussed above, they would also be more sensitive to possible material imperfections, absorption and losses, being based on a truly resonant phenomenon substantially different from the matching mechanism at the basis of the supercoupling.

Figure 7 also reports the distribution of the normal component of the electric field distribution (snapshot in time) along the waveguide at the supercoupling frequency $f_0$ for the geometry corresponding to the black line in Fig. 8a. Besides the energy squeezing and tunneling across the narrow channel, we notice that the phase is nearly uniform all over the channel. Despite its total length (the total length of the longer (outer) part of the channel is $36\,cm \simeq 1.8\lambda_0$, with $\lambda_0$ being the wavelength in free-space), the phase is transported in a static-like fashion from entrance to exit of the channel, despite abruptions in the E and H planes and bending. This is even more evident in the corresponding time-domain field animation, deposited in EPAPS [27].

For a smaller radius of curvature case (red line in Fig. 8a), Fig. 8b reports the geometry and the corresponding real part of the Poynting vector distribution on the bottom plane cross section. It is evident how the energy is dramatically squeezed in the ultranarrow channel and, despite the bending, the power is completely channeled towards the exit face at the supercoupling frequency.

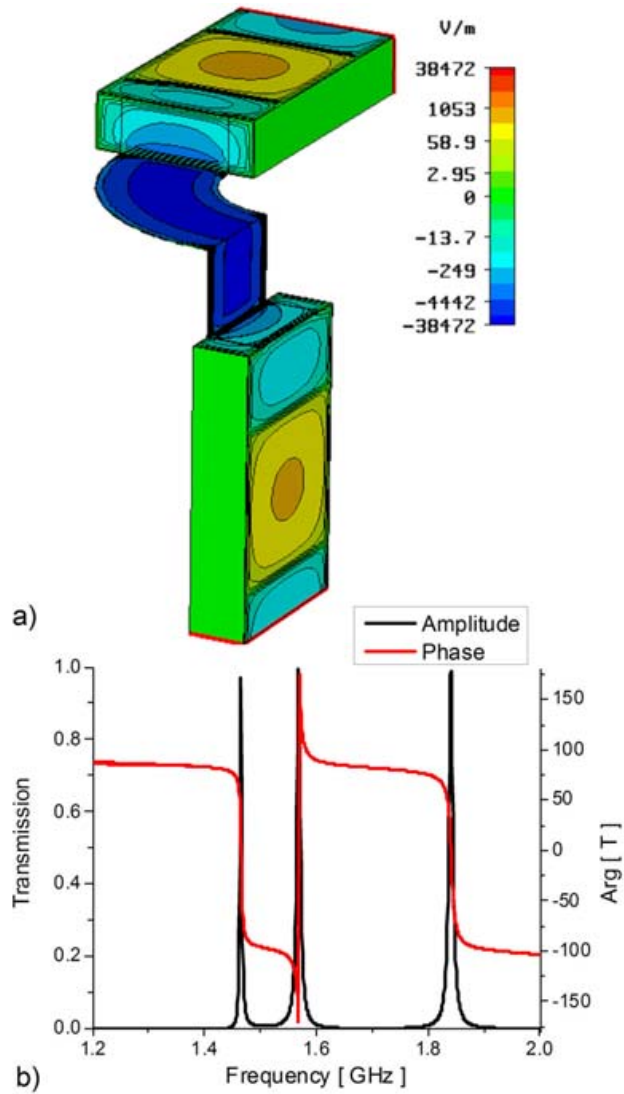

Figure 9 – (Color online) (a) Geometry and electric field distribution (snapshot in time, normal component) and (b) amplitude and phase of the transmission through a supercoupler, consistent with the geometry of Fig. 7, but with a 90° bend in the H plane following a 90° bend in the E plane.

As a final example, Fig. 9 reports the full-wave simulation results for a geometry in which the supercoupler channel in Fig. 7 is cascaded by another 90° bend, this time in the E plane. Despite the huge abruptions, width squeezing, bendings and manipulation of the channel geometry, consistent with the previous discussion, total transmission and zero phase delay are again achieved at the same ENZ frequency $f_0$. Once again, the whole geometry is filled with air and the ENZ operation is achieved by proper choice of the channel width. Field distribution (Fig. 9a) and transmission (Fig. 9b) are strikingly consistent with the previous discussions.

## 5. Conclusions

To conclude, we believe that the results presented in this extensive analysis may shed new light and insight onto the supercoupling effect, energy squeezing and anomalous tunneling effect produced by ENZ materials. Following our TL interpretation of this phenomenon, we have shown the drastic differences between this phenomenon and Fabry-Perot resonant tunneling and we have envisioned and presented novel alternative setups to achieve similar effects, involving hollow waveguides properly designed and connected. These results may be of importance for several applications, spanning waveguide connections and coupling, filtering, sensing, power conversion and absorption. Applying analogous concepts to plasmonic waveguides at cut-off, the possibility of extending these concepts at optical frequencies may be also envisioned.